# Magnetocaloric effect and nature of magnetic transition in nanoscale $Pr_{0.5}Ca_{0.5}MnO_3$

S. Narayana Jammalamadaka[1, 2*], S. S. Rao[3, 4,5,6*], S.V.Bhat[6], J. Vanacken[1], and V. V. Moshchalkov[1]

[1]*INPAC – Institute for Nanoscale Physics and Chemistry, Nanoscale Superconductivity and Magnetism, KU Leuven, Celestijnenlaan 200D, B–3001 Leuven, Belgium*

[2]*Department of Physics, Indian Institute of Technology Hyderabad, Ordnance Factory Estate Yeddumailaram, Andhra Pradesh, India, 502205.*

[3]*INPAC – Institute for Nanoscale Physics and Chemistry, Semiconductor Physics Laboratory, KU Leuven, Celestijnenlaan 200D, B–3001 Leuven, Belgium*

[4]*Materials Science Division, Army Research Office, Research Triangle Park, North Carolina 27709, USA*

[5]*Department of Material Science and Engineering, North Carolina State University, Raleigh, North Carolina 27695, USA*

[6]*Department of Physics, Indian Institute of Science, Bangalore – 560012, India*

*\* These authors contributed equally to this work*

## Abstract:

Systematic measurements pertinent to the magnetocaloric effect and nature of magnetic transition around the transition temperature are performed in the 10 nm $Pr_{0.5}Ca_{0.5}MnO_3$ nanoparticles (PCMO10). Maxwell's relation is employed to estimate the change in magnetic entropy. At Curie temperature ($T_C$) ~ 83.5 K, the change in magnetic entropy ($-\Delta S_M$) discloses a typical variation with a value 0.57 J/kg K, and is found to be magnetic field dependent. From the area under the curve ($\Delta S$ vs T), the refrigeration capacity (RC) is calculated at $T_C$ ~ 83.5 K and it is found to be 7.01 J/kg. Arrott plots infer that due to the competition between the ferromagnetic and anti-ferromagnetic interactions, the magnetic phase transition in PCMO10 is broadly spread over both in temperature as well as in magnetic field coordinates. Upon tuning the particle size, size distribution, morphology, and relative fraction of magnetic phases, it may be possible to enhance the magnetocalorific effect further in PCMO10.

Corresponding authors: surya@iith.ac.in , ssingam@ncsu.edu

## Introduction

The phenomenon of change in the temperature (heating or cooling) of a magnetic material with the applied magnetic field has been termed as magnetocaloric effect (MCE) and can be accomplished by magnetic entropy change ($\Delta$S) [1 - 4]. Magnetic materials exhibiting the change in temperature upon application of magnetic field have drawn widespread attention in the recent past due to their tremendous applications in the magnetic refrigeration technology. High efficiency, small volume, non–pollution, etc., are the appreciable advantages of magnetic refrigeration in comparison with the conventional gas refrigeration. Recent efforts in search of new magnetic refrigeration materials have extended the possible candidates to a broad series of compounds such as $CoFeO_4$ [5], $Pr_{0.65}(Ca_{0.7}Sr_{0.3})_{0.35}MnO_3$ [6], $La_{0.67}Sr_{0.33}MnO_{3-\delta}$ [7], Gd, $Gd_5(Si_2Ge_2)$ [4], $La_{0.8}Ca_{0.2}MnO_3$ [8], $Gd_{0.73}Dy_{0.27}$ [9], and La-Fe-Si [10]. Out of all, La-Fe-Si based compounds have been known to exhibit large MCE and could be accomplished by appropriately varying the ratio of various elements.

Among them, perovskite manganites having the general formula, $R_{1-x}A_xMnO_3$ (where R is a rare-earth ion and A an alkaline-earth ion) have been appropriate materials of choice owing to their fascinating properties like colossal magnetoresistance (CMR) and charge ordering (CO), to name a few [11 – 13]. These manganites have been considered as potential candidates for magnetic refrigeration based on MCE because of the large magnetic entropy change at moderate fields [14, 15]. Moreover, their extraordinary chemical stability and tunable phase transitions are also added advantages to choose manganites for the magnetic refrigeration applications.

$Pr_{0.5}Ca_{0.5}MnO_3$ has been one of the most appealing manganite, and in its bulk form it has been found to exhibit the charge, orbital orderings and charge exchange (CE) type anti-ferromagnetic insulating (AFM-I) phase at $T_{CO}$ = 245 K and $T_N$ = 175 K , respectively [16]. Suppression of CO phase and appearance of weak ferromagnetism at $T_C$ of 83.5 K have been reported in PCMO10 as shown in our previous work [17, 18]. In addition, ferromagnetic (FM) phase has been found to coexist with residual CO – AFM phase below spin – glass freezing temperature ($T_f$ = 37 K). Magnetic nanoparticles are more desirable than their bulk counterparts for an ideal magnetic refrigeration because the particle size distribution and interparticle interactions have been shown to broaden $\Delta S_M$ (T) over a wide temperature range, thus enhancing refrigeration capacity (RC). With that motivation, we put many efforts to realize MCE at $T_C$ in PCMO10. The present work demonstrates the MCE effect and nature of magnetic phase transition in PCMO10, which has not been reported to the best of our knowledge on this material.

**Experimental**

Preliminary results pertinent to structural and microstructural analysis of PCMO10 have already been published elsewhere [17, 18]. Nanoparticles of PCMO10 were wrapped up and inserted in non-magnetic teflon tape and straw respectively, in order to measure the isothermal magnetization measurements. Magnetization versus magnetic field (M-H) isotherms were collected in the temperature range of 5 - 100 K using a commercial vibrating sample magnetometer (VSM) (Oxford Instruments). At any given measuring temperature, the M-H curve was recorded after warming the sample to room temperature (280 K) so as to reduce the possible remnant effects.

**Results and discussion**

Extensive isothermal M-H (0-50 kOe) measurements were performed using VSM at several temperatures ranging from 5–100 K and are depicted in Figs. 1 (a) (b). At 5 K, a non-linear M-H curve is evident, hinting the presence of size induced FM phase, consistent with our earlier observations [17, 18]. However, as the temperature increases further, M-H curve exhibits linear character up to 100 K, inferring the gradual occurrence of paramagnetic (PM) phase due to the thermal energy which dominates the exchange energy.

Now we discuss and analyze the magnetization data to calculate the MCE, which would be induced *via* coupling of the magnetic sublattice with the external magnetic field. In general, the MCE can be realized by (a) magnetic entropy change (b) from the heat capacity measurements. Concerning the total entropy of the system, in general it would have three contributions (a) $S_{mag}$, the magnetic contribution, (b) $S_{lat}$, the lattice contribution and (c) $S_{el}$, the electronic contribution. Following the well-known procedure [19, 20], in the present article, the MCE is evaluated by considering only the magnetic entropy change ($-\Delta S_M$) and the Maxwell's equations are used as shown below. If we assume that $S_{lat}$ and $S_{el}$ are magnetic field independent.

$$\Delta S_M(T,H) = \int_0^{H_{max}} \left(\frac{\partial S}{\partial H}\right)_T dH \tag{1}$$

$$\left(\frac{\partial S}{\partial H}\right)_T = \left(\frac{\partial M}{\partial T}\right)_H \tag{2}$$

The magnetic entropy change can be rewritten as follows:

$$\Delta S_M(T,H) = \int\limits_{0}^{H_{max}} \left(\frac{\partial M}{\partial T}\right)_H dH \qquad (3)$$

From the isothermal M-H curves, the $-\Delta S_M$ is estimated in the temperature range of 5 – 100 K up to the magnetic field of 50 kOe using the equation (3). The thermal variation of $-\Delta S_M$ at different magnetic fields is shown in Fig. 2. Owing to the increase in the magnetic field to 50 kOe from 0 kOe, $\Delta S_M$ value of -0.57 J/kg-K is noticed at $T_C$ (83.5 K). From Fig. 2, it can be inferred that the $\Delta S_M$ (-0.320 J/kg-K) at spin-glass freezing temperature $T_f$ (37 K) is less in comparison with the value (-0.57 J/kg-K) at $T_C$. Earlier [6], in the case of $La_{0.125}Ca_{0.875}MnO_3$, low value of $-\Delta S_M$ (0.1 J/kg K) has been reported for 60 nm particles; $-\Delta S_M$ value of 0.255 J/kg K and 0.32 J/kg K has been reported for 60 nm particles of $La_{0.4}Ca_{0.6}MnO_3$ and for 32 nm particles of $La_{0.67}Sr_{0.33}MnO_3$ respectively [21, 22]. In a sharp contrast, yet, in another study [19], in a compound $La_{0.35}Pr_{0.275}Ca_{0.375}MnO_3$ of 50 nm sized particles, the $-\Delta S_M$ value as high as 6.2 J/kg K has been reported. We have also computed the refrigeration capacity (RC) using the equation

$$RC = \int \Delta S_M(T)dT.$$

For $\Delta H$ = 50 kOe, the computed RC for PCMO10 at $T_C$ is of 7. 015 J/kg, much smaller than the reported [19] value (225 J/kg K) for 50 nm particles of $La_{0.35}Pr_{0.275}Ca_{0.375}MnO_3$ at 50 kOe. The often quoted argument for such a large span in the values of $-\Delta S_M$ is intrinsic magnetic inhomogeneity and phase separation. However, a significant reduction in $\Delta S_M$ has routinely been obtained in the nanoparticle samples resulting mainly from the significant decrease of the magnetization upon size reduction and also on the distribution of particle size. Such explanation may not be true in the case of PCMO10 as it shows enhancement in the magnetization up on the size reduction down to nanoscale. Therefore, a further work is certainly needed to account for the observed low values of MCE.

To have deeper insights and gain further knowledge in realizing the amount of FM/AFM phase concentration, we have calculated the FM fraction from the isothermal magnetization (M-H) data. Following the well-known procedure [23], the FM fraction is estimated by subtracting the magnetization arising from the AFM component (linear part of the magnetization isotherms) from the saturated magnetization. The data is assembled in Fig. 3 as a function of temperature under various magnetic fields. From Fig. 3, it is evident that, FM fraction increases up to $T_f$. Further increase in temperature caused a decrease in FM fraction. With further increase in magnetic field, FM fraction grows as more and more spins start to align parallel to the external magnetic field. From Fig. 3, it is evident that maximum FM fraction is present at $T_f$. This may reveal that there could be a competition between coexisting FM and AFM phases, and the amount of FM fraction increases as we approach $T_f$. The maximum value of FM fraction at $T_f$ is found to be 55%. In one of the earlier theoretical works reported by Dong and co-authors [24] on spherical nanoparticles and cylindrical nanowires, the appearance of FM and the suppression of CO-AFM has been explained by proposing the core–shell model, where the nanoparticles such as PCMO10 have been assumed to compose a FM or partly FM shell with an AFM core. In their work [24], the authors have shown that the core contribution diminishes with the reduction in particle size. They have explained such an intriguing phenomenon by proposing the concept of 'surface phase separation' in which a growing FM tendency appears at the surface of nanoparticle/nanowire. It has been argued that the increase in the charge density due to the unscreened Coulomb interactions changes the surface layer from AFM/CO to phase separated electronic state with a FM tendency on the surface [24]. We believe that such a phenomenon in PCMO10 may yield a FM fraction of about 55% at $T_f$. The observed FM fraction (55%) is comparable with the values that have been reported for other nanoscale manganites. For example, the FM fraction in $La_{0.5}Ca_{0.5}MnO_3$ nanoparticles (25 nm) is 40% at 5 K [25], in $La_{0.5}Ca_{0.5}MnO_3$ nanoparticles (15 nm) it is 91% at 10 K [26], and in $Ca_{0.82}La_{0.18}MnO_3$ nanoparticles (20 – 30 nm) it is about 22% at 5 K [27].

**Nature of magnetic phase transition**

Extensive research [28 - 31] has been carried out to probe the nature of magnetic phase transitions and the behaviour of critical exponents both in micro crystalline as well as in nanocrystalline CMR manganites. For example, in detailed and systematic studies [30, 31], it has been concluded that size reduction down to nanoscale leads to a crossover from first order to second order magnetic phase transition in CMR manganites such as $La_{0.67}Ca_{0.33}MnO_3$. As we discussed in the previous sections, size reduction in CO bulk manganites such as PCMO induces ferromagnetism. Here, our aim is to probe the nature of such FM transition using Arrott plots. We believe that this is the first attempt of its kind in exploring the nature of magnetic phase transition induced by size reduction in nanoscale CO manganites such as PCMO10. The nature of the magnetic transition can be obtained from the curvature [32] of isotherm plots of $M^2$ vs. H/M, M being the experimentally observed magnetization and H the magnetic field. Such phenomenon has been widely used to determine the order of the magnetic phase transition. A positive or a negative curvature of the experimental H/M vs. $M^2$ curve indicates a second order or a first order transition, respectively.

The magnetization isotherms for PCMO10 drawn in the form of Arrott plots are shown in Fig. 4. The isotherms were recorded for the temperature range 5 -100 K. The Arrott plots $M^2$ vs H/M displayed in Fig. 4 exhibit the positive curvature of the curves at all temperatures studied. This indicates that the transition between FM and PM phases is of second order in nature. The Arrott plots did not allow us to determine a critical temperature of magnetic transition since the extensions of the plots do not reach a center of the $M^2$ and H/M coordinate system. This may suggest us that the magnetic transition is broadly spread both in temperature and in magnetic field coordinates, due to a competition between FM and AFM interactions. The broadening effect may be additionally enhanced by the presence of local magnetic inhomogeneities of PCMO10.

**Conclusion and Outlook**

The competition between the ferromagnetic and anti-ferromagnetic phases may result in a frustrated magnetic state and may further lead to the complex magnetic behavior in 10 nm $Pr_{0.5}Ca_{0.5}MnO_3$ particles. The small particle size, size distribution and second order phase transition may cause for the apparent low values of magnetocaloric effect (0.57 J/Kg–K) and refrigeration capacity (7.015 J/Kg). A maximum value of FM fraction 55% is observed at the spin–glass freezing temperature, and might be accounted from the core–shell model. From the Arrott plots, it may be inferred that due to the competition between the ferromagnetic and anti-ferromagnetic interactions, the magnetic transition is broadly spread both in temperature and in magnetic field coordinates. By appropriately engineering the particle size, its distribution, sample morphology, and relative percentage of magnetic fractions one can enhance the magneto calorific effect and may consider employing the current material for magnetic cooling applications. Another important parameter that may play a big role would be the type of phase transition. If one can determine the ways to obtain the first order phase transition in this material, there would indeed be a good chance to enhance the magnetocalorific effect.

**Acknowledgements**

SNJ would like to thank KU Leuven, for research fellowship, SSR acknowledges CSIR, Government of India for fellowship. SVB thanks DST, India for project funding under NSTI. This work is supported by the K U Leuven Excellence financing (INPAC), by the Flemish Methusalem financing and by the IAP network of the Belgian Government.

**References:**

1. M. Tishin, in Handbook of Magnetic Materials, edited by K. H. J. Buschow (North Holland, Amsterdam, 1999), Vol. **12**, pp. 395 - 524.

2. V. K. Pecharsky and K. A. Gschneidner, Jr., J. Magn. Magn. Mater. **200**, 44(1999).

3. H. Takeya, V. K. Pecharsky, K. A. Gschneidner Jr., J. O. Mooman, Appl. Phys. Lett. **64**, 2739 (1994).

4. K. A. Gschneidner Jr., V. K. Pecharsky, J. Appl. Phys. **85**, 5365 (1999).

5. E. V. Gopalan, I. A. Al-Omari, D. S. Kumar, Y. Yoshida, P. A. Joy and M. R. Anantharaman, Appl. Phys. A **99**, 497 (2010)

6. Biswas, T. Samanta, S. Banerjee, and I. Das, Appl. Phys. Lett. **92**, 212502 (2008).

7. W. J. Lu, X. Luo, C. Y. Hao, W. H. Song, and Y. P. Sun, J. Appl. Phys. **104**, 113908 (2008).

8. Z. B. Guo, Y. W. Du, J. S. Zhu, H. Huang, W. P. Ding, D. Feng, Phys. Rev. Lett. **78**, 1142 (1997).

9. V. K. Pecharsky, K. A. Gschneidner, Appl. Phys. Lett. **70**, 3299 (1997).

10. L. Jia, J. R. Sun, J. Shen, Q. Y. Dong, J. D. Zou, B. Gao, T. Y. Zhao, H. W. Zhang, F. X. Hu, and B. G. Shen, J. Appl. Phys. **105**, 07A924 (2009).

11. For a review see Y. Tomioka and Y. Tokura, in *Colossal Magnetoresistive Oxides*, edited by Tokura _Gordon and Breach, New York, (2000).

12. Asamitsu, Y. Tomioka, H. Kuwahara and Y. Tokura, Nature (London) **388**, 50 (1997).

13. P. Wagner, I. Gordon, L. Trappeniers, J. Vanacken, F. Herlach, V. V. Moshchalkov, and Y. Bruynseraede, Phys. Rev. Lett., **81**, 3980 (1998); K. Miyano, T. Tanaka, Y. Tomioka and Y. Tokura, Phys. Rev. Lett., **78**, 4257 (1997).

14. S. Nair and A. Banerjee, J. Phys.: Condens. Matter **16**, 8335 (2004).

15. P. Chen, Y. W. Du and G. Ni, Europhys. Lett., **52,** 589 (2000).

16. Y. Tomioka, A. Asamitsu, H. Kuwahara, Y. Moritomo, and Y. Tokura, Phys. Rev. B **53**, R1689 (1996).

17. S. S. Rao and S. V. Bhat  J. Phys. Condens. Matter **22**, 116004 (2010); S. S. Rao, K. N. Anuradha, S. Sarangi, and S. V. Bhat, Appl. Phys. Lett*., * **87**, 182503 (2005); S. S. Rao, S. Tripathi, D. Pandey, and S. V. Bhat, Phys. Rev. B **74**, 144416 (2006), and references therein.

18. S. Narayana Jammalamadaka, S. S. Rao, J. Vanacken, A. Stesmans, S. V. Bhat and V. V. Moshchalkov, AIP ADVANCES **1**, 042151 (2011); S. Narayana Jammalamadaka, S. S. Rao, J. Vanacken, S. V. Bhat and V. V. Moshchalkov AIP ADVANCES **2**, 012169 (2012).

19. M. H. Phan, S. Chandra, N. S. Bingham, H. Srikanth, C. L. Zhang, S. W. Cheong, T. D. Hoang and H. D. Chinh, Appl. Phys. Lett., **97**, 242506 (2010).

20. G. J. Liu, J. R. Sun,  J. Z. Wang, and B. G. Shen, Appl. Phys. Lett.,  **89**, 222503 (2006); R. Venkatesh, M. Pattabiraman, S. Angappane, G. Rangarajan, K. Sethupathi, K. Jessy, M. Feciouru-Morariu, R. M. Ghadimi, and G. Guntherodt Phys. Rev. B **75,** 224415, (2007).

21. C. L. Lu, K. F. Wang, S. Dong, J. G. Wan, J.-M. Liu, and Z. F. Ren, J. Appl. Phys **103**, 07F714 (2008).

22. W. J. Lu, X. Luo, C. Y. Hao, W. H. Song and Y. P. Sun, J. Appl. Phys  **104**, 113908 (2008).

23. A. M. Gomes, F. Garcia, A. P. Guimarães, M. S. Reis and V. S. Amaral, Appl. Phys. Lett., **85**, 4974 (2004).

24. S. Dong, F. Gao, Z. Q. Wang, J. M. Liu, and Z. F. Ren, Appl. Phys. Lett., **90**, 082508 (2007). S. Dong, R. Yu, S. Yunoki, J. M. Liu, and E. Dagotto, Phys. Rev. B **78**, 064414 (2008). A. Rostamnejadi, M. Venkatesan, J. Alaria, M. Boese, P. Kameli, H. Salamati and J. M. D. Coey J. Appl. Phys. **110**, 043905 (2011).

25. Z. Jirák, E. Hadová, O. Kaman, K. Knížek, M. Maryško, E. Pollert, M. Dlouhá and S. Vratislav, Phys. Rev. B  **81**, 024403 (2010).

26. T. Sarkar, B. Ghosh, A. K. Raychaudhuri, and T. Chatterji, Phys. Rev. B **77**, 235112 (2008).

27. Y. Wang, and H. J. Fan, Appl. Phys. Lett., **98**, 142502 (2011).

28. B. Padmanabhan, H. L. Bhat, Suja Elizabeth, Sahana Rößler, U. K. Rößler, K. Dörr, and K. H. Müller, Phys. Rev. B **75,**  024419 (2009).

29. Jiyu Fan, L. Ling, Bo Hong, Lei Zhang, Li Pi, and Y. Zhang, Phys. Rev. B **81,** 144426 (2010).

30. R. N. Bhowmik, A. Poddar, R. Ranganathan, and C. Mazumdar, J. Appl. Phys. **105**, 113909 (2009)

31. V. Markovicha, I. Fitab,c, D. Mogilyanskyd, A. Wisniewskib, R. Puzniakb, L. Titelmand, L. Vradmand, M. Herskowitze, G. Gorodetskya, Superlattices and Microstructures **44,** 476 (2008).

32. S. K. Banerjee, Phys. Lett., **12**, 16 (1964).

**Figure captions**

**Fig. 1:** Magnetization isotherms of PCMO10 at different temperatures in the field range 0 – 50 kOe.

**Fig. 2:** Temperature dependence of magnetic entropy change $-\Delta S_M$ at different applied fields up to 50 kOe for the PCMO10.

**Fig. 3:** Variation of ferromagnetic fraction with the temperature at different magnetic fields.

**Fig. 4:** $M^2$ vs.(H/M) plots for isotherms of magnetization for PCMO10. The curvature is always positive at any measured temperature, denoting the second-order character of ferromagnetic to paramagnetic transition at $T_C = 83.5$ K.

**Figures**

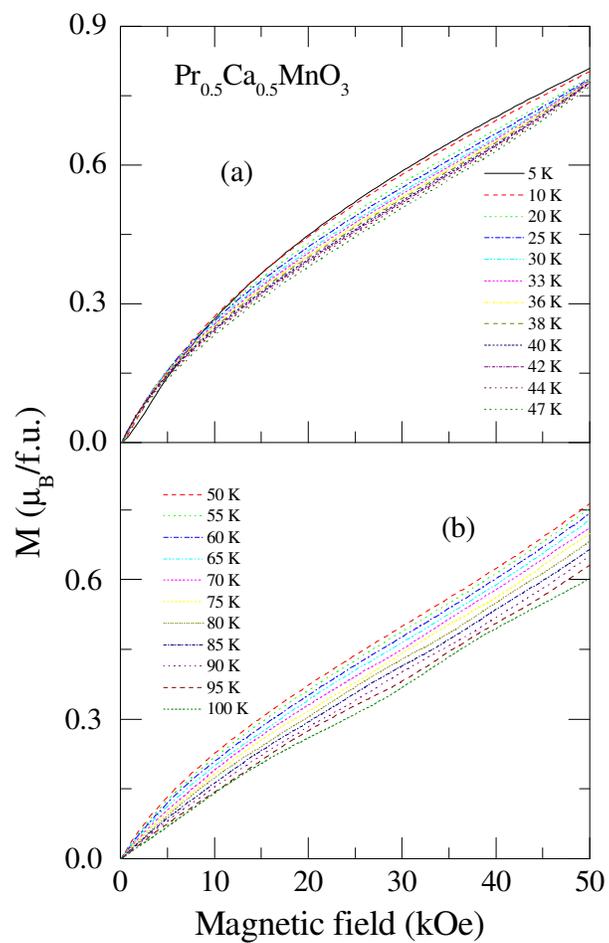

**Fig. 1:** Magnetization isotherms of PCMO10 at different temperatures in the field range 0–50 kOe.

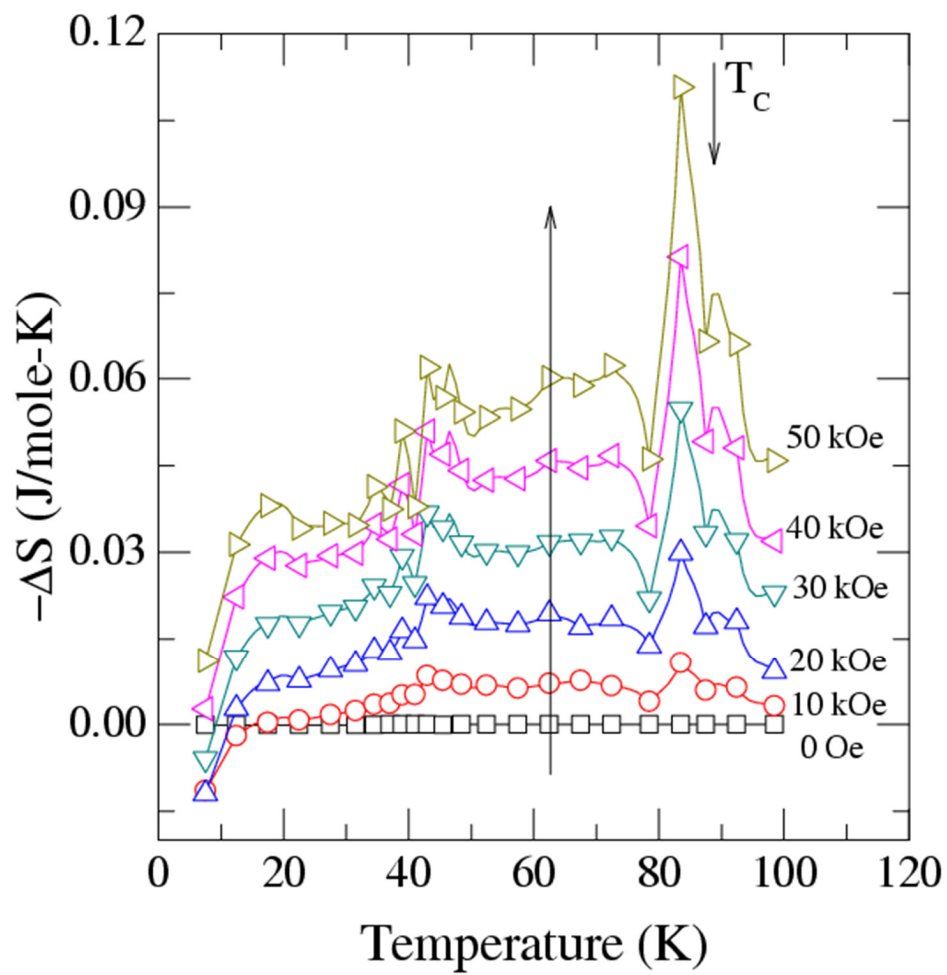

**Fig. 2:** Temperature dependence of magnetic entropy change -$\Delta S_M$ at different applied fields up to 50 kOe for the PCMO10.

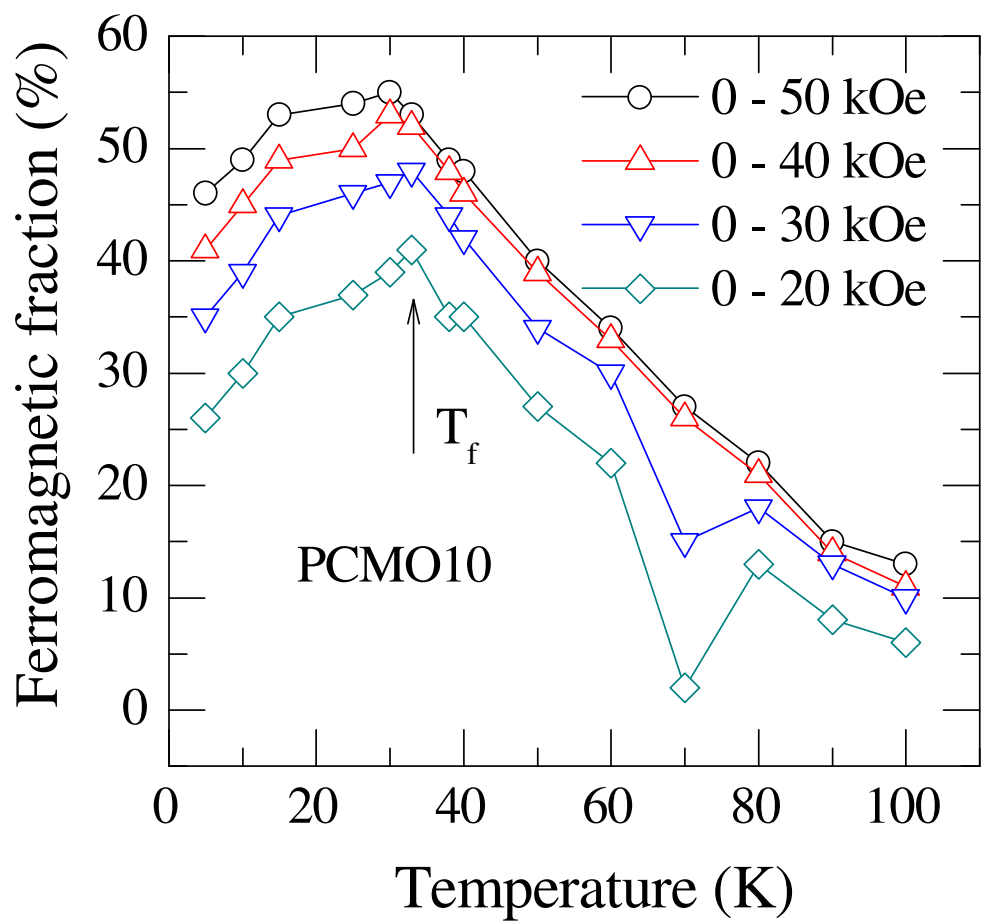

**Fig. 3:** Variation of ferromagnetic fraction with the temperature at different magnetic fields.

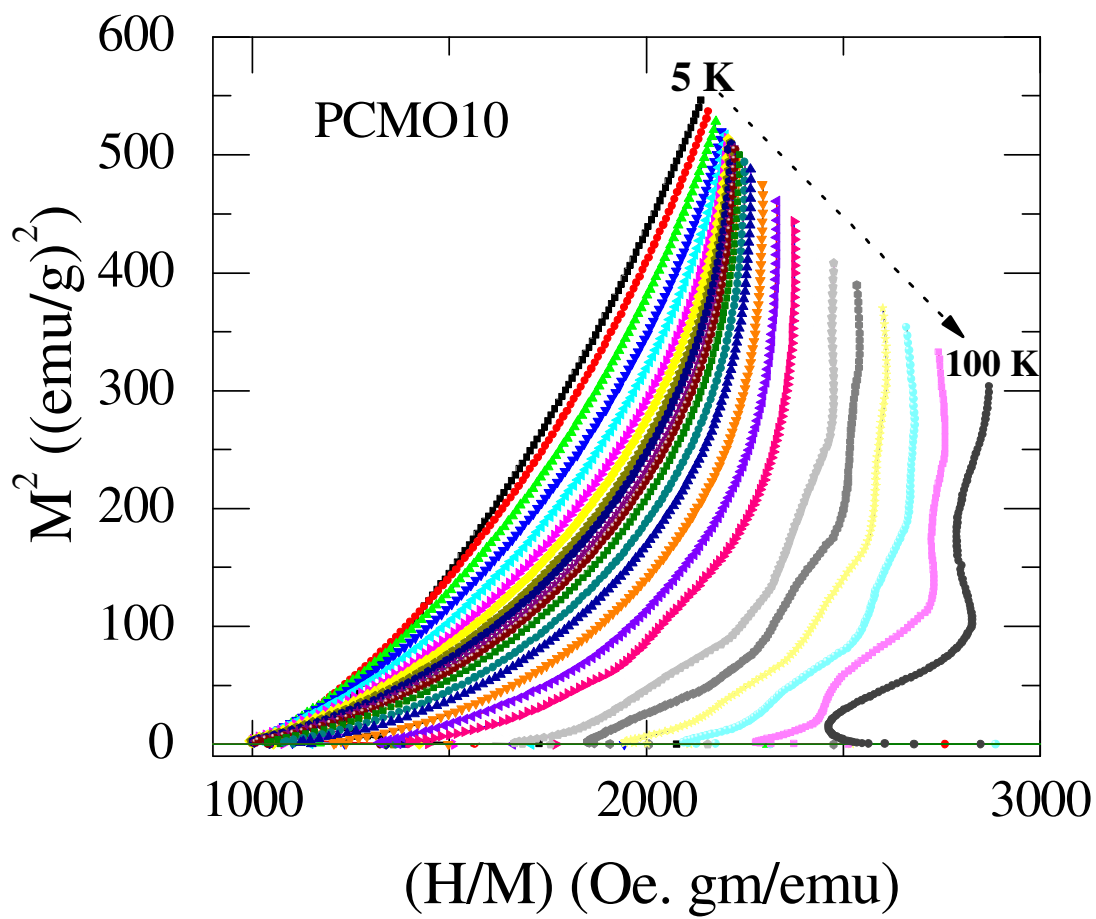

**Fig. 4:** $M^2$ vs.(H/M) plots for isotherms of magnetization for PCMO10. The curvature is always positive at any measured temperature, denoting the second-order character of ferromagnetic to paramagnetic transition at $T_C = 83.5$ K.